\documentclass[aps,pra,preprint,superscriptaddress,longbibliography]{revtex4-1}

\usepackage{amsmath}
\usepackage{amsthm}
\usepackage{amsfonts}
\usepackage{amssymb}
\usepackage{xcolor,graphicx}
\usepackage{tikz}
\usepackage{color}
\usepackage[pdftex]{hyperref} 
\usepackage{longtable}
\usepackage{array}
\usepackage{dsfont}

\begin{document}
	
	\title{Heat flow reversal in a trapped-ion simulator}
	
	\author{P.~U. Medina Gonz\'alez}
	\affiliation{AMC-Tecnologico de Monterrey, Escuela de Ingenier\'ia y Ciencias, Ave. Eugenio Garza Sada 2501, Monterrey, N.L., Mexico, 64849}
	
	\author{I. Ramos Prieto}
	\email[e-mail: ]{iran@inaoep.mx}
	\affiliation{Instituto Nacional de Astrof\'{i}sica, \'{O}ptica y Electr\'{o}nica, Calle Luis Enrique Erro No. 1. Sta. Ma. Tonantzintla, Pue. C.P. 72840, Mexico}

	\author{B. M. Rodr\'iguez-Lara}
	\email[e-mail: ]{bmlara@tec.mx}
	\affiliation{Tecnologico de Monterrey, Escuela de Ingenier\'ia y Ciencias, Ave. Eugenio Garza Sada 2501, Monterrey, N.L., Mexico, 64849}
	\affiliation{Instituto Nacional de Astrof\'{i}sica, \'{O}ptica y Electr\'{o}nica, Calle Luis Enrique Erro No. 1. Sta. Ma. Tonantzintla, Pue. C.P. 72840, Mexico}
	
	\date{\today}
	
	\begin{abstract}
	We propose a trapped-ion platform to simulate a reconfigurable spin-spin Hamiltonian related to quantum thermodynamic processes. 
	Starting from an experimental model describing two trapped-ions under slightly off-resonant first sideband driving with individually controlled driving phases, we follow an operational quantum optics approach to show that it produces an effective model appearing in recent quantum thermodynamics proposals.
	We show that projection into the vibrational ground state manifold allows full analytic treatment.
	As a practical example, we take experimental data from a $^{187}$Yb$^{+}$ trap and numerically simulate the reversal of heat flow  between two thermal spins controlled by their quantum correlations.
	\end{abstract}
	
	
	\maketitle

\section{Introduction}

Thermodynamics played an essential role in the development of quantum theory \cite{Einstein1905p132}. 
In recent decades, quantum mechanics is helping redefine thermodynamics providing a quantum origin for its laws \cite{Jarzynski1997p2690,Jarzynski1997p5018,BookDeffner2019}.
In this sense, for example, the three level maser helped introduce the idea of quantum heat engines to study the theory of heat for Carnot cycles \cite{Scovil1959p262,Geusic1967p343}. 
In these engines, the working substance is a quantum system; for example, harmonic oscillators \cite{Lin2003p046105,Wang2015p062134}, two level atoms \cite{Feldmann2000p4774,Quan2007p031105,Altintas2014p032102}, and systems from quantum electrodynamics like cavities \cite{Quan2006p036122}, trapped ions \cite{Abah2012p203006,Chand2017p032111,Chand2018p052147}, or optomechanical systems \cite{Zhang2014p023819}. 
A fundamental concept in this framework is the energy transfer from one quantum system to another that is closely related to the variation of quantum correlations \cite{Uzdin2015p031044,Klatzow2019p110601}.

Quantum correlations play a fundamental role in describing the exchange of energy in bipartite quantum systems, in order to test the second law of thermodynamics \cite{Brandao2008p873,Henao2018p062105,Park2017,Ma2019p062303}. 
According to Clausius, \textit{heat can never pass from a colder to a warmer body without some other change, connected therewith, occurring at the same time} \cite{Clausius1854p481}.
At the microscopic level, contemporaneous results show that quantum correlations can help reversing the natural flow of heat transfer, understood as the exchange of energy \cite{Henao2018p062105}.
Furthermore, this inversion can be achieved without reversing the arrow of time \cite{Latune2019p033097}.
It is possible to think about two basic approaches to realize quantum thermodynamics; its full physical realization in a multipartite open quantum system \cite{Alicki1979pL103,Kosloff1984p1625,Chen2010p063002,Kosloff2013p2100,delCampo2013p100502,Muga2019p045001} or its quantum simulation in a highly controllable platform that is robust to decoherence.
A recent example of the experimental quantum simulation of heat engines using the nuclear magnetic resonance platform \cite{Micadei2019p2456} focuses on the the reversal of heat flow due to correlations in the initial state of the system. 
Here, we will explore ion-traps, another well-established platform for quantum simulation \cite{Blatt2012p277}, and build upon experimental proposals to realize integer-spin chains \cite{Senko2015p021026} to show that it is possible to use them as a reconfigurable platform for the simulation of quantum heat engines.

We focus on a trap where two ions interact with two common vibrational modes of their center of mass motion and require off-resonant first sideband driving with independent phase control. 
In the following section, we analytically show we can take the model into an integer-spin chain form with individual control of the effective magnetic fields and spin-spin interactions. Then, we show that our model reduces to a form related to proposals for the simulation of quantum heat engines and the reversal of the heat flow direction in the vibrational ground state manifold.
We focus on the latter and use experimental data from an $^{187}$Yb$^{+}$ trap to demonstrate its experimental feasibility using our platform. 
Finally, we close with a brief summary and conclusion.

\section{Trapped ion platform}

We start from the experimental realization of a quantum integer-spin chain with controllable interactions in trapped ions \cite{Senko2015p021026},
\begin{equation}\label{H}
\hat{H}=\hat{H}_0 + \sum_{j} \hbar \Omega_{j} \left\{e^{ \mathrm{i}\sum_{m}\eta_{jm}\left(\hat{a}_m^\dagger+\hat{a}_m\right) } e^{\mathrm{i}(\phi_j-\omega t)}+  e^{ -\mathrm{i}\sum_{m}\eta_{jm}\left(\hat{a}_m^\dagger+\hat{a}_m\right) } e^{-\mathrm{i}(\phi_j-\omega t)} \right\}\left(\hat{\sigma}_+^{(j)}+\hat{\sigma}_-^{(j)}\right),
\end{equation}
where we allow for controllable phase differences between driving lasers. 
The free Hamiltonian,  
\begin{equation}\label{H_0}
\frac{\hat{H}_{0}}{\hbar}=\sum_{j} \frac{\omega_e}{2}\hat{\sigma}_z^{(j)} + \sum_{m} \omega_m \hat{n}_m,
\end{equation}
describes two internal levels of two trapped-ions in terms of Pauli matrices, $\hat{\sigma}_{i}^{(j)}$ with $i = z, \pm$ and $j=1,2$, and two common vibration modes of their center of mass motion in terms of the phonon creation (annihilation) operators, $\hat{a}^{\dagger}_{m}$ ($\hat{a}_{m}$) with $m=1,2$ and phonon number operator $\hat{n}_{m} = \hat{a}^{\dagger}_{m} \hat{a}_{m}$.
The frequencies $\omega_{e}$,  $\omega$, $\omega_{m}$, are those of the two-level system energy gap, the driving fields, which we assume identical in both cases, and the $m$-th common vibrational mode, in that order.
We assume full and independent control of the coupling strength $\Omega_{j}$ between internal levels and vibrational modes, and the phase of driving lasers $\phi_{m}$, leading to a Lamb-Dicke factor $\eta_{jm}$ between the $j$-th ion and the $m$-th mode.

Carefully extending a standard theoretical procedure for single ion systems \cite{Leibfried2003p281}, we move into the frame defined by the free Hamiltonian $\hat{H}_{0}$ and perform a rotating wave approximation,
\begin{align}
\begin{split}
&\frac{\hat{H}_{\mathrm{RWA}}}{\hbar} =\mathrm{i}\sum_{j} \Omega_je^{-\frac{\eta_{j1}^2+\eta_{j2}^2}{2}}  \Bigg\{\left[\eta_{j1}F(\eta_{j1},\hat{n}_1;\eta_{j2},\hat{n}_2)\hat{a}_1e^{\mathrm{i}(\delta-\omega_1)t}+\eta_{j2}G(\eta_{j1},\hat{n}_1;\eta_{j2},\hat{n}_2)\hat{a}_2e^{\mathrm{i}(\delta-\omega_2)t}\right] \\
&\times\hat{\sigma}_+^{(j)}e^{\mathrm{i}\phi_j} + \left[\eta_{j1}\hat{a}_1^\dagger F(\eta_{j1},\hat{n}_1;\eta_{j1},\hat{n}_2)e^{-\mathrm{i}(\delta-\omega_1)t}+\eta_{j2}\hat{a}_2^\dagger G(\eta_{j1},\hat{n}_1;\eta_{j2},\hat{n}_2)e^{-\mathrm{i}(\delta-\omega_2)t}\right] \hat{\sigma}_{-}^{(j)}e^{-\mathrm{i}\phi_j}\Bigg\},
\end{split}
\end{align}
where we neglect terms including the frequency $\Delta = \omega_{e} + \omega$ and keep those with frequency $\delta = \omega_{e} - \omega$.
This is done in the laboratory by choosing a driving frequency $\omega$ such that $\Delta \gg \delta \sim \omega_{m}$.
The effective coupling constants,
\begin{align}
\begin{split}
F(\eta_{j1},\hat{n}_1;\eta_{j2},\hat{n}_2)&= {_1}F_{1}(-\hat{n}_1;2;{\eta_{j1}}^2)L_{\hat{n}_2}({\eta_{j2}}^2),\\
G(\eta_{j1},\hat{n}_1;\eta_{j2},\hat{n}_2)&= {_1}F_{1}(-\hat{n}_2;2;{\eta_{j2}}^2)L_{\hat{n}_1}({\eta_{j1}}^2),
\end{split}
\end{align}
are given in terms of Kummer confluent hypergeometric function $_{1}F_{1}(a; b; z)$ and the generalized Laguerre polynomial $L_{a}^{(b)}(z)$ \cite{BookOlver2010}. 
In the Lamb-Dicke regime $\eta_{m,j}\langle\hat{a}_m+\hat{a}_m^\dagger\rangle^{1/2}\ll1$, we can approximate these functions to yield the following expression,
\begin{align} 
\begin{split} \label{eq:HLD}
\frac{\hat{H}_{\mathrm{LD}}}{\hbar}  \simeq & \sum_{j} \frac{\delta}{2} \hat{\sigma}_{z}^{(j)}  + \sum_{m} \omega_{m} \hat{n}_{m} + 
 \mathrm{i}\sum_{j,m} \Omega_j\eta_{jm}\left(\hat{a}_m\hat{\sigma}_+^{(j)}e^{\mathrm{i}\phi_j}-\hat{a}_m^\dagger\hat{\sigma}_-^{(j)}e^{-\mathrm{i}\phi_j} \right),
\end{split}
\end{align}
once we move into the rotating frame defined by the first two terms in the right hand side.
Now, we use a small rotation \cite{Klimov2000p063802},
\begin{align}
\hat{\mathcal{R}}=\exp\left[-\mathrm{i}\sum_{l,s=1,2}g_{ls}\left(\hat{a}_l\hat{\sigma}_+^{(s)}e^{\mathrm{i}\theta_s}+\hat{a}_l^\dagger\hat{\sigma}_-^{(s)} e^{-\mathrm{i}\theta_s}\right)\right],
\end{align}
with parameters, 
\begin{equation}
g_{mj}=\frac{\Omega_j\eta_{mj}}{\delta-\omega_m},
\end{equation}
to realize that the effective Hamiltonian of the system is that of an integer spin chain,
\begin{align}
\begin{split}
\frac{\hat{H}_{R}}{\hbar} \simeq&\sum_{m}\omega_m\hat{n}_m+\sum_{j,m}V_{jm}\left(2\hat{n}_m+1\right)\hat{\sigma}_z^{(j)} +  \sum_{j,m\neq l}\frac{\Omega_j^2\eta_{jm}\eta_{jl}}{\delta-\omega_l}\left(\hat{a}_l^\dagger\hat{a}_m+\hat{a}_l\hat{a}_m^\dagger\right)\hat{\sigma}_z^{(j)} + \\
&\sum_{j\neq s}J_{js}\left(\hat{\sigma}_+^{(j)}\hat{\sigma}_-^{(s)}e^{\mathrm{i}\Delta\phi_{js}}+\hat{\sigma}_-^{(j)}\hat{\sigma}_+^{(s)}e^{-\mathrm{i}\Delta\phi_{js}}\right),
\end{split}
\end{align}
with tunable interactions, 
\begin{equation}
V_{jm}=\frac{\Omega_j\eta_{jm}}{\delta-\omega_m} \quad\mbox{and}\quad J_{js}=\Omega_j\Omega_s\sum_{m}\frac{\eta_{jm}\eta_{sm}}{\delta-\omega_m},
\end{equation}
as long as the experiment parameters fulfil $\delta \sim \omega_{m}$ and $ \Omega_{j} \eta_{jk}  \ll (\delta - \omega_{m})$ in the Lamb-Dicke regime.
The first right hand side term of this effective Hamiltonian describes the free common vibrational modes, the second term has the form of Stark effect between the phonon field and the two-level systems, the third term is a  Stark-like effect but in terms of a beam-splitter-like interaction between the common vibrational modes. 
The fifth term is our objective, a spin-spin interaction controllable by the relative phase difference between the driving fields $\Delta\phi_{js}=\phi_j-\phi_s$. 
This model is our first contribution, a first principles algebraic derivation of a trapped-ion platform for the simulation of quantum thermodynamics based on integer-spin chain models.
In the following, we will provide a working example.

\section{Evolution in the vibrational ground state manifold}

For the sake of simplicity, we consider the common vibrational modes initialized in their ground state. 
In the Lamb-Dicke regime, the coupling between internal states and vibrational modes is small and the number of excitations remains unchanged. 
Thus, we can project our model into the vibrational ground state and recover an effective spin chain model, 
\begin{equation}
\begin{split}
\frac{\hat{\mathcal{H}}_{S}}{\hbar} = & \sum_{j,m}V_{jm}\hat{\sigma}_z^{(j)} + 
\sum_{j\neq s}J_{js}\left(\hat{\sigma}_+^{(j)}\hat{\sigma}_-^{(s)}e^{\mathrm{i}\Delta\phi_{js}}+\hat{\sigma}_-^{(j)}\hat{\sigma}_+^{(s)}e^{-\mathrm{i}\Delta\phi_{js}}\right),
\end{split}
\end{equation}
with individual control of the magnetic fields driving each spin and the spin-spin interaction. 
This model defines our simulation frame as it is the basic building block for spin-spin proposals that realize quantum heat engines~\cite{Chand2017p032111} and reversal of the heat flow direction using quantum correlations~\cite{Micadei2019p2456}.
We focus on the latter. 
Thus, we consider an initial state in the laboratory frame,
\begin{equation}
\hat{\rho}(0) = \hat{\rho}_{s}(0) \otimes \vert 00 \rangle \langle 00 \vert,
\end{equation}
where the ions are initialized in separable thermal states
plus an arbitrary correlation,
\begin{align}
\begin{split}
\hat{\rho}_s(0) =& \frac{e^{-\Gamma_{+}}}{\mathcal{Z}_1\mathcal{Z}_2}\begin{pmatrix}
e^{\Gamma_{+}}&0&0&0\\
0&e^{\Gamma_-}&\alpha&0\\
0&\alpha^*&e^{-\Gamma_-}&0\\
0&0&0&e^{-\Gamma_{+}}\\
\end{pmatrix},
\end{split}
\end{align}
where the ions pseudo-temperatures appear via $\Gamma_{j} = \hbar\omega_{e} \beta_{j} / 2$ with $\beta_{j} = (k_{B} T_{j})^{-1}$ and Boltzmann constant $k_{B}$ and $\Gamma_{\pm} = \Gamma_{1} \pm \Gamma_{2}$.
The partition functions are $\mathcal{Z}_j=\mathrm{Tr}[e^{-\beta_j\hat{\mathcal{H}}_j}]$ with $\hat{\mathcal{H}}_j=\frac{\hbar\omega_e}{2}(1-\hat{\sigma}_z^{j})$.
As each ion is prepared in a state equivalent to a local Gibbs state $\hat{\rho}_{T_{j}}^{(j)}=e^{-\beta_j\hat{\mathcal{H}}_j}/\mathcal{Z}_j$, the energy exchange between them can be interpreted as heat transfer \cite{Jarzynski2004p230602}.
The correlation parameter $\alpha$ must fulfil $|\alpha|^2 \leq 1/2 - \left(1+\tanh \Gamma_1 \right)^2 \left(1+\tanh \Gamma_2 \right)^2 / 8$ in order to keep the trace of the spin reduced density matrix equal to one.

Choosing driving phases $\phi_{1m} = \phi_{2m} \equiv \phi_{m}$ allows us to construct an analytic evolution operator,
\begin{equation}
\hat{\mathcal{U}}(t)=\begin{pmatrix}
\mathcal{U}_{11}(t)&0&0&0\\
0&\mathcal{U}_{22}(t)&\mathcal{U}_{23}(t)&0\\
0&- \mathcal{U}_{23}^{\ast}(t)&\mathcal{U}_{22}^{\ast}(t)&0\\
0&0&0&\mathcal{U}_{11}^{\ast}(t)
\end{pmatrix}
\end{equation}
with matrix components, $ \mathcal{U}_{11}(t) = e^{-\mathrm{i}V_{+} t}$, $\mathcal{U}_{22}(t) = \cos \Omega t - \mathrm{i} (V_{-}/\Omega) \sin \Omega t$ , $\mathcal{U}_{23}(t) = - \mathrm{i}(J/\Omega) e^{\mathrm{i} \Delta \phi}\sin \Omega t$, defined in terms of the auxiliary frequencies $V_{\pm} =\sum_{m} (V_{1m} \pm V_{2m})$ and $\Omega = \sqrt{ V_{-}^{2} + J^{2} } $ with $J = J_{12} + J_{21}$ and $\Delta \phi = \phi_{1} - \phi_{2}$.

It is straightforward but cumbersome to evolve the initial state in the laboratory frame.
The fact that population inversion measurements, 
\begin{align}
\mathrm{Tr} \left[ \hat{\rho}(t) \hat{\sigma}_{z}^{(j)} \right] \simeq \mathrm{Tr} \left[ \hat{\mathcal{U}}^{\dagger}(t) \hat{\rho}(0) \hat{\mathcal{U}}(t) \hat{\sigma}_{z}^{(j)} \right],
\end{align}
are similar in the laboratory and simulation frames due to the transformations and approximations used, makes it a natural choice to work with the pseudo-energies for each simulated spin, 
\begin{eqnarray}
\mathcal{Q}_{j} = \frac{1}{2} \hbar \omega_{e} \mathrm{Tr} \left[ \hat{\rho}(t) \hat{\sigma}_{z}^{(j)} \right],
\end{eqnarray} 
as they can be reconstructed from the fluorescence of each ion in the laboratory, 
\begin{flalign}
\mathcal{Q}_1(t) = \frac{\hbar \omega_e}{2\Omega^2} \bigg[& J^2 \sin^2 \Omega t \left(   \tanh \Gamma_1 - \tanh  \Gamma_2  - \frac{4 V_- r}{J} \cos \Delta \theta \right) + \\ \notag
&~\Omega^2 \left(1 - \tanh  \Gamma_1 - \frac{2 J r}{\Omega} \sin \Delta \theta \sin 2 \Omega t\right)\bigg], \\
\mathcal{Q}_2(t) = \frac{\hbar \omega_e}{2\Omega^2} \bigg[& J^2 \sin^2 \Omega t \left(   \tanh  \Gamma_2 - \tanh  \Gamma_1  + \frac{4 V_- r}{J} \cos \Delta \theta \right) + \\ \notag
&~\Omega^2 \left( 1 - \tanh  \Gamma_2  + \frac{2 J r}{\Omega} \sin \Delta \theta \sin 2 \Omega t\right) \bigg],
\end{flalign}
where we use a polar decomposition of the complex correlation parameter  $\alpha=r e^{\mathrm{i} \theta}$ and define $\Delta \theta \equiv \Delta \phi - \theta$.
We calculate a pseudo-energy difference,
\begin{flalign}
\mathcal{Q}_{12}(t) = -\frac{\hbar \omega_{e}}{ \Omega^2} \bigg[ &  \frac{1}{2} \left( \tanh \Gamma_{1} - \tanh \Gamma_{2} \right) \left( J^2 \cos 2 \Omega t + V_-^2 \right) + \\ \notag
&~2 J r \left( 2 V_{-} \cos \Delta \theta \sin^{2} \Omega t + \Omega  \sin \Delta \theta  \sin 2 \Omega  t \right) \bigg], 
\end{flalign}
and its flux from the first to the second ion,
\begin{flalign}
\frac{d}{dt}\mathcal{Q}_{12}(t) = \frac{\hbar \omega_{e}}{ \Omega} \bigg[&  J^2  \left( \tanh \Gamma_{1} - \tanh \Gamma_{2} \right)  \sin 2 \Omega t  -  \\ \notag
 &~4 J r \left(  V_{-} \cos \Delta \theta \sin 2 \Omega t + \Omega  \sin \Delta \theta  \cos 2 \Omega  t \right) ].
\end{flalign}
We can already try to reverse the direction of the pseudo-heat-flow by adequately choosing the amplitude $r$ and phase $\theta$ of the correlation parameter $\alpha$.
In the following, we focus on parameters from a particular experiment that make it simpler to realize this.

\section{Heat flow direction reversal}

Let us consider the experimental parameters from an actual Ytterbium ion trap \cite{Senko2015p021026}, $\{\omega_{0}, \omega_{1}, \omega_{2} \} /2\pi = \{ 12.643\,\mathrm{GHz}, 3.5838\,\mathrm{MHz}, 3.5305\,\mathrm{MHz} \}$ tuned to provide coupling strengths $\Omega_{1}/2\pi= \Omega_{2}/2\pi = 300\,\mathrm{KHz}$ and Lamb-Dicke parameters $\eta_{jm} = 0.049$. 
We ask for slightly asymmetric off-resonant first sideband driving such that $\delta = 3.5571\,\mathrm{MHz} < (\omega_{1} + \omega_{2})/2 = 3.55715\,\mathrm{MHz}$ and a phase difference $\Delta \phi_{j} = -\pi/2$ between driving fields.
We focus on closed dynamics as Ytterbium traps report up to one second coherence times~\cite{Mount2013p093018} and parameters provide $J_{12} = J_{21} = J/2 = 191.17\,\mathrm{Hz}$, $V_{+} = 382.34167\,\mathrm{Hz}$, $V_{-} = 0$, and $\Omega = J$ such that we recover a simplified pseudo-heat-flow,
\begin{align}
	\begin{split}
	\frac{d}{dt}\mathcal{Q}_{12}(t) =&  \hbar \omega_{e} J \left[  \left( \tanh \Gamma_{1} - \tanh \Gamma_{2} \right)  \sin 2 J t - 4 r \sin \Delta \theta  \cos 2 J  t   \right],
	\end{split}
\end{align}
providing a viable simulation in the millisecond range.

\begin{figure}[ht!]
	\centering
	\includegraphics{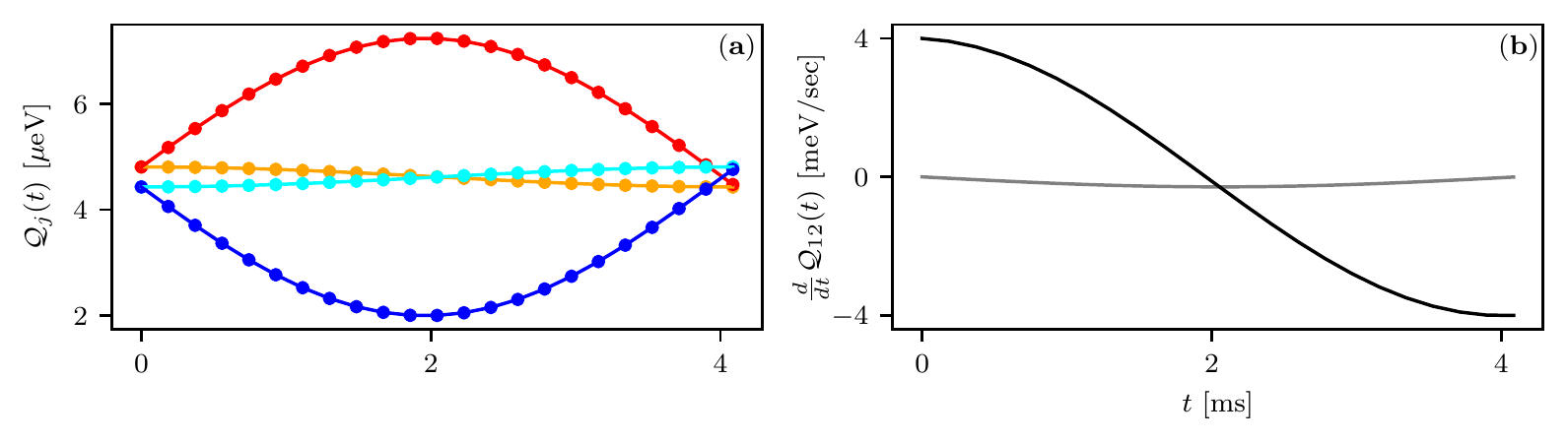}
	\caption{(a) Pseudo-energies of the simulated spins with initial correlation $\alpha=0$ in light red and blue, $\alpha = 0.05$ in red and blue, and (b) their corresponding pseudo-heat-flow in gray for $\alpha=0$ and black for $\alpha=0.05$. Dots show the full numerical evolution under dynamics provided by $\hat{H}_{LD}$ within a vibrational manifold with up to 10 excitations.}\label{fig:Fig1}
\end{figure}

Absence of correlations and pseudo-temperature difference, yields a null pseudo-heat-flow.
For identical pseudo-temperatures, the direction of the flow depends on the phase difference between the driving fields and that of the correlation $\Delta \theta$: a zero value produces null flow, a negative (positive) one induces flow from the first (second) to the second (first) simulated spin.
For no correlations and pseudo-temperatures $T_{1} > T_{2}$, the flow immediately becomes positive; that is, a decreasing $T_{1}$ and increasing $T_{2}$ as expected from standard thermodynamics, see the pale red and blue lines in Fig.~\ref{fig:Fig1} for ions prepared in initial thermal states with $T_{1} = 265~\mathrm{mK}$ and $T_{2} = 255~\mathrm{K}$.
In this case, it is possible to calculate a correlation value that produces the opposite effect, see the strong blue and red lines in Fig.~\ref{fig:Fig1} for $\alpha = r e^{i \theta} = 0.05 $ such that $\Delta \theta = \Delta \phi = \pi/2$.
It is important to stress that the quantum simulation is valid in the range $2 \Omega t \in [0, \pi/2]$.

\begin{figure}[ht!]
	\centering
	\includegraphics{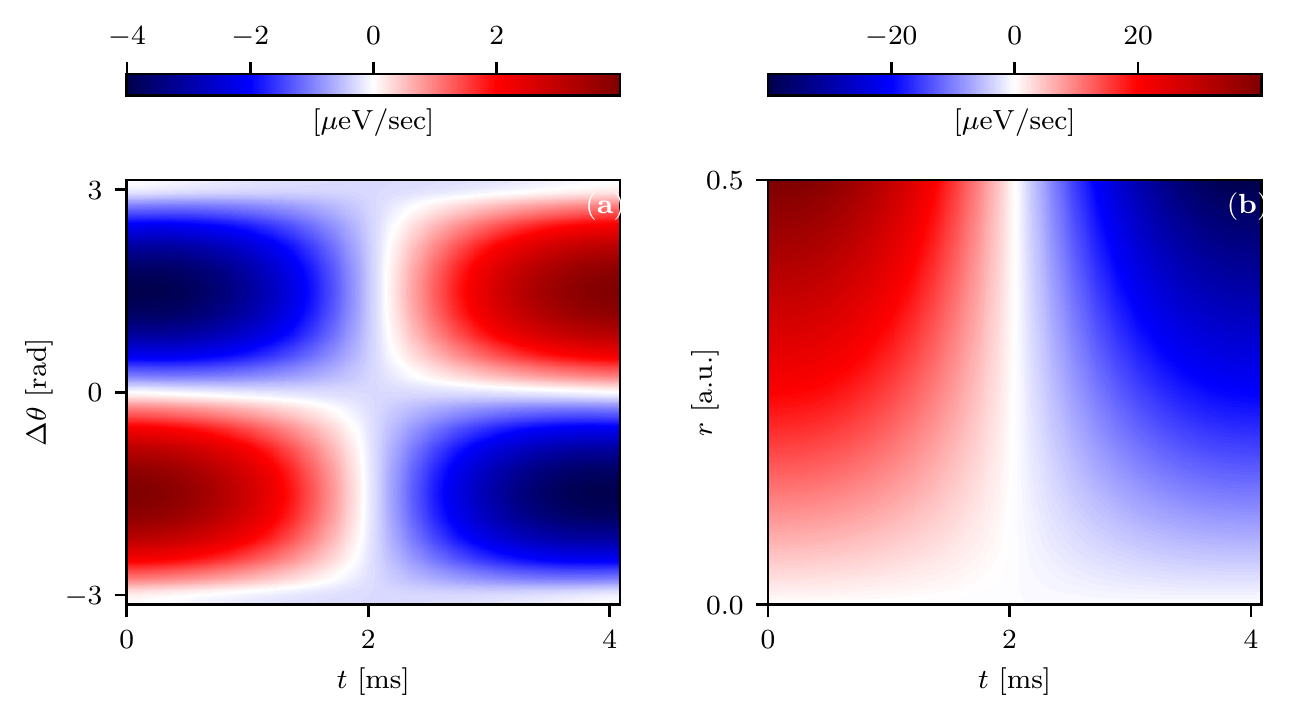}
	\caption{Pseudo-heat-flow evolution for (a) fixed correlation amplitude $r=0.05$ and variable phase difference $\Delta \theta \in [-\pi, \pi]$, and (b) vice-versa for $\Delta \theta = - \pi/2$ and $r \in [0, 0.5]$.}\label{fig:Fig2}
\end{figure}

In general for a given experimental parameter set, we decide where to place ourselves in the simulation by choosing a suitable value of the correlation amplitude $r$ for a fixed phase difference, Fig.~\ref{fig:Fig2}(a), or vice-versa, Fig.~\ref{fig:Fig2}(b).

\section{Conclusion}

We used a first principles model and an operator approach to show that two trapped-ions provide a reconfigurable platform for the simulation of quantum thermodynamics. 
Our proposal requires off-resonant first sideband driving and the ability to control the phase differences between driving lasers.
It simulates an integer-spin chain where both the magnetic fields driving each spin and the spin-spin interactions are controlled individually.
Particular simplifications of our model are related to current theoretical proposals on the field of quantum thermodynamics.

As a practical example, we took experimental data from an $^{187}$Yb$^{+}$ trap to show that it is possible to simulate the reversal of heat flow direction in this platform.
In our simulation, the internal levels are initialized into thermal states plus a correlation and the vibrational modes to the ground state. 
The population difference of individual ion defines pseudo-energies and  the time derivative of their difference provides a pseudo-heat-flow whose direction is controlled by the correlation.

We believe that our proposal in its most general form can play a significant role for the simulation of complex quantum thermodynamics processes in trapped-ions.

\begin{acknowledgments}
P.~U.~M.~G. acknowledges financial support from AMC under the Summer Research Program 2019. 
I.~R.~P. and B.~M.~R.-L. acknowledge financial support from CONACYT under project CB-2015-01/255230.
B.~M.~R.-L. acknowledges fruitful discussion with Benjamin Jaramillo \'Avila.
\end{acknowledgments}


%

\end{document}